\documentstyle[psfig]{ioplppt}

\def\expval#1{{\langle #1 \rangle}}
\begin{document}
\jl{3}
\letter{Internal energy and condensate fraction of a trapped
interacting Bose gas}
\author{A Minguzzi, S Conti and M P Tosi}
\address{Istituto Nazionale di Fisica della Materia and Classe di
Scienze, Scuola Normale Superiore, Piazza dei Cavalieri, 
I-56126 Pisa, Italy}
\begin{abstract}
We present a semiclassical two--fluid model for an interacting
Bose gas confined in an anisotropic harmonic trap and solve it in the
experimentally relevant region for a spin--polarized gas of
${}^{87}$Rb atoms, obtaining the temperature dependence of the
internal energy and of the condensate fraction. Our results are in
agreement with recent experimental observations by Ensher \etal.
\end{abstract}
\vskip1cm
Bose--Einstein condensation (BEC) has recently been realized in dilute
vapours of spin--polarized alkali atoms, using advanced techniques for
cooling and trapping\cite{anderson95,hulet,ketterle1,ketterle2,ensher96}. These
condensates consist of several 
thousands to several million atoms confined in a well which is
generated from nonuniform magnetic fields. The confining potential is
accurately harmonic along the three Cartesian directions and has
cylindrical symmetry in most experimental setups.

The determination of thermodynamic properties such as the condensate
fraction and the internal energy as functions of temperature is at
present of primary interest in the study of these
condensates\cite{ketterle2,ensher96}. The nature of BEC is
fundamentally affected by the 
presence of the confining potential\cite{Groot50} and finite size
corrections are appreciable, leading for instance to a reduction in
the critical temperature\cite{grossmann95,haugerud96,minguzzi96,Kirsten96}. 
Interaction effects are very small in the normal phase but become
significant with the condensation--induced density increase.
The correction to the transition temperature
due to interactions has been recently computed by Giorgini
\etal\cite{giorgini96}. 

The temperature dependence of the condensate fraction was recently
measured\cite{ensher96} for a sample of around 40000 ${}^{87}$Rb atoms,
the observed lowering in transition temperature being in agreement
with theoretical predictions within experimental resolution. In the
same work the internal energy was measured during ballistic expansion
and found to be significantly higher in the BEC phase than 
predicted by the ideal--gas model.
While the increase is easily understood as a consequence
of the interatomic repulsions, a quantitative estimate is still lacking. 

In this work we present a two--fluid mean--field model which is able to
explain the above--mentioned effects, giving results in agreement with
experiment for both the condensate fraction and the internal energy as
functions of temperature.  

We describe the condensate by means of the Gross--Pitaevskii (GP)
equation for its wave function $\Psi(r)$,
\begin{equation}\label{eqGP}
-{\hbar^2\nabla^2\over 2m}\Psi(r) + V^{\mathrm{ext}}(r)\Psi(r)+2gn_1(r)\Psi(r) + g \Psi^3(r) = \mu\Psi(r)\end{equation}
where $g=4\pi\hbar^2 a/m$, $a$ being the scattering length,
$V^{\mathrm{ext}}({\bf r})=m \omega^2 
(x^2+y^2+\lambda^2 z^2)/2$ is the confining potential and $n_1(r)$ is
the average non--condensed particle distribution. The factor $2$ in
the third term arises from exchange\cite{griffin96} and
we neglect the 
term involving the off--diagonal density of non--condensed particles. 
Following Bagnato \etal\cite{bagnato87} we treat the
non--condensed particles as non--interacting 
bosons in an effective potential $V^{\mathrm{eff}}(r)=V^{\mathrm{ext}}(r)+2g n_1(r) + 2g
\Psi^2(r)$.
Thermal averages are computed with a standard semiclassical Bose--Einstein
distribution in chemical equilibrium with the condensate,
i.e. at the same chemical potential $\mu$. In particular, 
the density $n_1(r)$ is
\begin{eqnarray}
n_1(r)&=&{1\over (2\pi\hbar)^3} \int  {d^3p\over \exp\left\{\left({p^2\over
2m} + V^{\mathrm{eff}}(r) - \mu\right)/k_BT\right\} -1}\nonumber\\
&=&{( m k_B T)^{3/2}\over (2\pi)^{3/2}\hbar^3} \sum_{j\ge 1} {1\over j^{3/2}}
\exp\left\{-j(V^{\mathrm{eff}}(r)-\mu)/k_BT\right\}
\label{eqserien1}\,.\end{eqnarray}
We fix the chemical potential from the total number of particles $N$
\begin{equation}\label{eqntot}
N = N_0  + \int {\rho(E) dE \over
\exp\left\{\left(E - \mu\right)/k_BT\right\} -1}\end{equation}
where $N_0=\int \Psi^2(r) d^3r$ and the semiclassical density of
states is 
\begin{equation}\label{eqrhodef}
\rho(E)={(2m)^{3/2}\over4\pi^2\hbar^3} \int_{V^{\mathrm{eff}}(r)<E}
\sqrt{E-V^{\mathrm{eff}}(r)} d^3r\,. \end{equation}
This completes the self--consistent closure of the model.

Equation (\ref{eqGP}) can be solved analytically in the
experimentally relevant situation $N\sim 10^4\div 10^5$ and $a/a_\perp\sim
10^{-2}$, where $a_\perp=\sqrt{\hbar/m\omega}$.
Except for a small region close 
to the phase transition the interaction parameter $N_0 a/a_\perp$ entering
the GP equation is large and the kinetic energy can be neglected. 
This yields
\begin{equation}\label{eqpsir}
\Psi^2(r)={\mu-V^{\mathrm{ext}}(r)-2gn_1(r)\over g} 
\theta(\mu-V^{\mathrm{ext}}(r)-2gn_1(r))
\end{equation}
where $\theta(x)=0$ ($1$) for $x<0$ ($x>0$). 
The present strong--coupling solution neglects the condensate
zero--point energy $E_0^c=\hbar\omega(1+\lambda/2)$. 
As Giorgini, Pitaevskii and Stringari\cite{giorgini96} pointed out
finite--size effects are thereby excluded.

Before presenting the complete numerical solution of the
self--consistent model defined by 
equations (\ref{eqserien1})--(\ref{eqpsir}) and comparing its
predictions with existing experimental
data\cite{ensher96}, we display perturbative solutions at 
zero-- and first--order.

An approximate semi--analytical solution can be obtained by treating
perturbatively interactions involving the ``dilute gas'' of non--condensed
particles. To zero order in $gn_1(r)$ we have 
\begin{equation}\label{eqn0mu}
N_0=\left(2\mu\over\hbar\omega\right)^{5/2} {a_\perp\over
15\lambda a}\end{equation}  
and equation (\ref{eqrhodef}) gives 
\begin{equation}
\fl
\rho_0(E)={1\over\pi\lambda (\hbar\omega)^3}\left[
2\sqrt\mu\left(E\!-\!\mu\right)^{3/2} + 
 E^2 \mbox{arctn}{\sqrt{E\!-\!\mu\over \mu}} 
+(2\mu\!-\!E)^2 \ln {\sqrt{|E\!-\!2\mu|}\over\sqrt\mu\!+\!\sqrt{E\!-\!\mu}}
\right] \label{eqrhoc}
\end{equation}
for $\mu>0$ and
\begin{equation}\label{eqn0sopratc}
\rho_0(E)={E^2\over 2\lambda (\hbar\omega)^3}\end{equation}
for $\mu<0$.
The self--consistent zero--order solution is then completed by equation
(\ref{eqntot}). We remark that no
assumption of weak interactions {\em within the condensate} has been made.

We now proceed to compute the first order correction to the above
zero--order solution.
We take 
\begin{equation}\label{eqn0approx}
\Psi^2(r) = {\mu - V^{\mathrm{ext}}(r) - 2 gn_1(0)\over g}\theta(\mu -
V^{\mathrm{ext}}(r) - 2 gn_1(0))\end{equation}
and expand equation (\ref{eqrhodef}) to first order in
$g[n_1(r)-n_1(0)]$.
The choice of the expansion parameter $g[n_1(r)-n_1(0)]$  ensures that
the perturbative expansion is regular, since the
correction to $\rho$ vanishes where the zero--order term
vanishes. With the additional approximation $g\Psi^2(r)\simeq \mu\theta(\mu)$
in the first--order term we get
\begin{equation}\label{eqrhotutta}
\rho(E)=\rho_0(E\!-\!2gn_1(0)) + \delta\rho_1
(E\!-\!2gn_1(0)-\mu\theta(\mu))\end{equation}  
where 
\begin{equation}\label{eqrho1e}\fl
\delta\rho_1(E)={4\over\sqrt{2\pi}\lambda}
{a\over a_\perp} \left(k_B T\over\hbar\omega\right)^{3/2}
\sum_{j\ge1} {E
e^{j\mu\theta(-\mu)/k_BT}\over j^{3/2}(\hbar\omega)^2} 
\left[1-{}_1F_1\left({3\over2}, 2, -j{E\over k_BT}\right)\right]\end{equation}
${}_1F_1$ being the Kummer confluent hypergeometric function.
The self--consistent first--order solution is then completed by equation
(\ref{eqntot}). 

\begin{figure}[tb]
\centerline{\psfig{figure=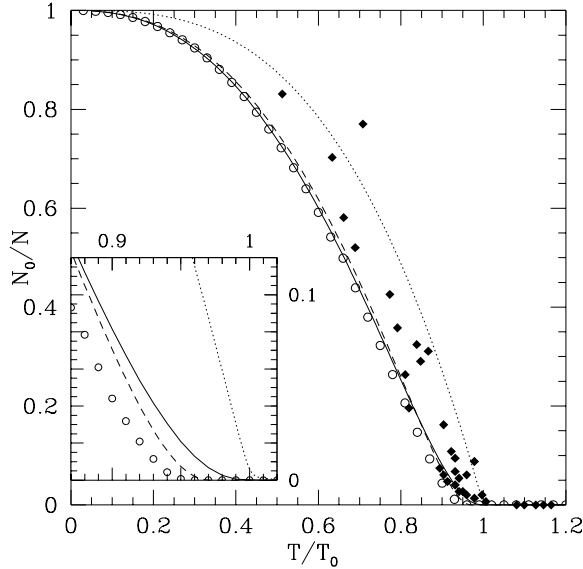,width=0.6\textwidth}}
\caption{Condensate fraction 
obtained in the two--fluid model compared with the experimental data  of Ensher
\etal\protect\cite{ensher96} (diamonds) and with the ideal gas result
(dotted curve). We present results obtained from the zero--order
solution (full curve), from the first--order perturbative treatment
(dashed curve) and from the complete numerical solution (circles). 
The inset
is an enlargement of the region around $T_{\mathrm c}$.}
\label{fign0}
\end{figure}

\begin{figure}[tb]
\centerline{\psfig{figure=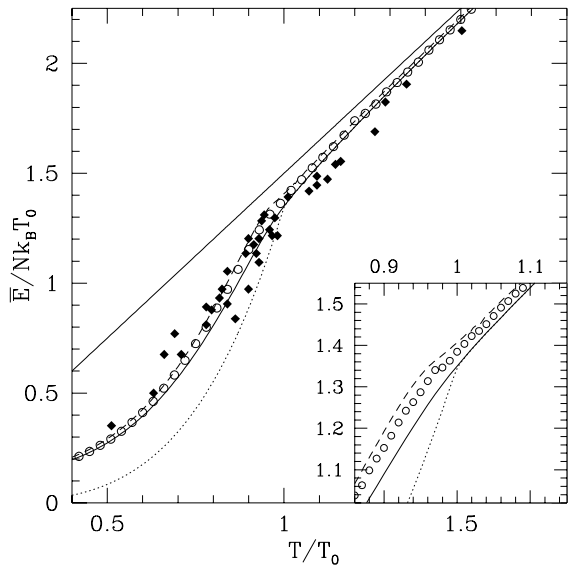,width=0.6\textwidth}}
\caption{Sum of kinetic and interaction energy, as defined in the text,
obtained in the two--fluid model compared with the experimental data  of Ensher
\etal\protect\cite{ensher96} (diamonds) and with the ideal gas result
(dotted curve). We present results obtained from the zero--order
solution (full curve), from the first--order perturbative treatment
(dashed curve) and from the complete numerical solution (circles). 
The straight line is the classical Maxwell--Boltzmann result. The inset
is an enlargement of the region around $T_{\mathrm c}$.}
\label{figeqst}
\end{figure}

\begin{figure}[tb]
\centerline{\psfig{figure=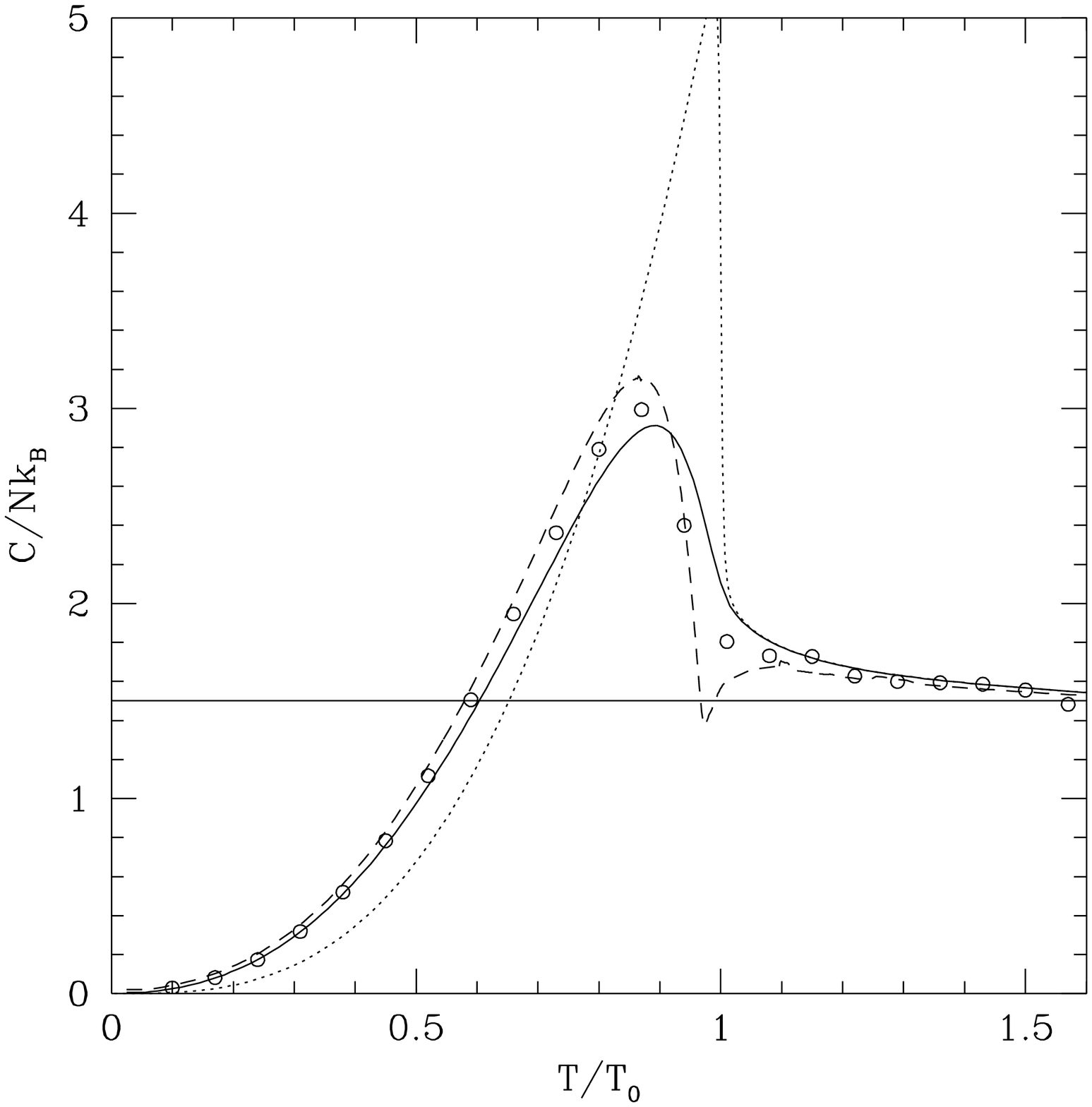,width=0.6\textwidth}}
\caption{Specific heat $C=d\bar E/dT$ 
obtained in the two--fluid model compared with the ideal gas result
(dotted curve). We present results obtained from the zero--order
solution (full curve), from the first--order perturbative treatment
(dashed curve) and from the complete numerical solution (circles). 
The straight line is the classical Maxwell--Boltzmann result.
The anomalous behaviour obtained from the first--order perturbative
solution near $T_c$ is an artifact.}
\label{figcv}
\end{figure}

We have solved numerically the simplified two--fluid model, first treating the
parameter $gn_1(0)$ to all orders (equations
\ref{eqserien1}--\ref{eqpsir}), then to zero order (equations 
\ref{eqntot} and \ref{eqn0mu}--\ref{eqn0sopratc})
and finally to first order (equations \ref{eqntot} and
\ref{eqn0approx}--\ref{eqrho1e}). Each case involves solving the integral
equation (\ref{eqntot}) to obtain $\mu$ as a function of $N$ and $T$;
the non--perturbative solution also involves the local
nonlinear problem posed by equations (\ref{eqserien1}) and (\ref{eqpsir}). 
The small differences between our three results
justify {\it a posteriori} a perturbative treatment.
Experimental parameters are taken from the work of Ensher \etal
\cite{ensher96}:  
$\lambda=\sqrt8$, $N=40000$ and $a/a_\perp=0.0062$. 
We have verified numerically that our results depend 
weakly on $N$ in the region explored in the experiments and therefore
have used a fixed $N=40000$ in all our computations. 
We use as energy units the semiclassical ideal--gas critical temperature 
$k_BT_0=\hbar\omega (N\lambda/\zeta(3))^{1/3}$, $\zeta(3)\simeq 1.202$
being the Riemann zeta function. 

Figure \ref{fign0} compares the temperature dependence of the condensate
fraction $N_0/N$  with the experimental results of
Ensher \etal\cite{ensher96}. Lowering of the transition temperature
due to interactions is clearly visible, even if the smoothness of our
results around the transition prevents a precise assessment of an
interaction--induced shift in $T_{\mathrm c}$ from the numerical solution. 
It should be noticed that the
strong--coupling solution of the GP equation is not
valid for $T$ close to $T_{\mathrm c}$, since it requires $N_0 \gg a_\perp/a
\simeq 160$, and that our mean field model does not
include critical fluctuations. Both effects being relevant only in a
narrow window around $T_{\mathrm c}$\cite{giorgini96}, we expect our results 
to be meaningful in most of the temperature range. 
Recently Giorgini \etal\cite{giorgini96} solved
numerically the Popov 
approximation to the finite--temperature generalization of the
GP equation within a semiclassical WKB approximation. 
Their results for the temperature dependence of the condensate
fraction are in very good agreement with the predictions of our
more naive model except for $|T-T_{\mathrm c}|/T_{\mathrm c}<0.05$,
where they find a sharp change in the slope of $N_0(T)$. Their result
for the interaction--induced shift in critical temperature $\delta
T_{\mathrm c}/T_{\mathrm c} \simeq -1.33 N^{1/6} a/a_\perp\simeq
-0.048$ is also in good agreement with our curves.

Figure \ref{figeqst} reports our results for the temperature dependence
of the internal energy. We remark that 
the experimentally measured quantity is the sum of the kinetic energy
and of the interaction energy, not including the confinement potential
energy due to the rapid switching off of the trapping
potential\cite{BaymPethick96}.
The average single particle energy $\expval{E}_{\mathrm{nc}}=\int
E \rho(E)dE / \left\{\exp\left[(E-\mu)/k_BT\right]-1\right\} $
obtained from the semiclassical density of states 
contains twice the interaction energy, and -- assuming that on average
the kinetic and potential terms are equal -- is twice the measured
quantity. 
The kinetic energy of condensed atoms is negligible in our strong--coupling
limit and their interaction energy per particle is
$\expval{E}_{\mathrm{c}}={1\over2} g \int \Psi^4(r) d^3r$. 
The quantity directly comparable to the experimental data is therefore
$\bar E=(\expval{E}_{\mathrm{nc}}(N-N_0)/2 +
\expval{E}_{\mathrm{c}})/N$, which we plot  
in figure \ref{figeqst} obtaining good agreement with the measured
values.
The calculated internal energy does not contain any sharp feature at
transition, paralleling the result discussed above for the condensate
fraction. Correspondingly the rapid rise in the specific heat is
considerably smoothed with respect to the ideal--gas result (see
figure \ref{figcv}).  
Apart from this small region around transition, 
our results on $\bar E(T)$ above and below $T_{\mathrm c}$ imply a
significant reduction of the increase in specific heat across the
phase transition.

In conclusion, we have presented a mean--field, semiclassical two--fluid
model and discussed its perturbative and non--perturbative solution in
the experimentally 
relevant parameter range. Our results on the temperature dependence
of the condensate fraction and of the internal energy are in 
agreement with recent experimental measurements, accounting for the
pronounced increase in internal energy with respect to the noninteracting boson
case measured below $T_{\mathrm c}$. We have also verified that
our model reproduces the results obtained for the condensate fraction
with a more refined theory by Giorgini \etal.

\section*{Acknowledgements}

We thank Dr E.~A.~Cornell for making his data available to us prior to
publication


\section*{References}

\begin{thebibliography}{10}

\bibitem{anderson95}
Anderson~M~H, Hensher~J~R, Matthews~M~R, Wieman~C~E and Cornell~E~A 1995  {\em
  Science} {\bf 269}  198

\bibitem{hulet}
Bradley~C~C, Sackett~C~A, Tollett~J~J and Hulet~R~G 1995  {\em Phys. Rev.
  Lett.} {\bf 75}  1687

\bibitem{ketterle1}
Davis~K~B, Mewes~M~O, Andrews~M~R, van Druten~N~J, Durfee~D~S, Kurn~D~M and
  Ketterle~W 1995  {\em Phys. Rev. Lett.} {\bf 75}  3969

\bibitem{ketterle2}
Mewes~M~O, Andrews~M~R, van Druten~N~J, Kurn~D~M, Durfee~D~S and Ketterle~W
  1996  {\em Phys. Rev. Lett.} {\bf 77}  416

\bibitem{ensher96}
Ensher~J~R, Jin~D~S, Matthews~M~R, Wieman~C~E and Cornell~E~A 1996
{\em Phys. Rev. Lett.} {\bf 77} 4984

\bibitem{Groot50}
{De Groot}~S~R, Hooyman~G~J and {ten Seldam}~C~A 1950  {\em Proc. Roy. Soc.
  London A} {\bf 203}  266

\bibitem{grossmann95}
Grossmann~S and Holthaus~M 1995  {\em Phys. Lett. A} {\bf 208}  188

\bibitem{haugerud96}
Haugerud~H, Haugset~T and Ravndal~F preprint cond-mat/9605100 (unpublished)

\bibitem{minguzzi96}
Minguzzi~A, Chiofalo~M~L and Tosi~M~P 1997 {\em N. Cimento D} (to appear)

\bibitem{Kirsten96}
Kirsten~K and Toms~D~J 1996 {\em Phys. Lett.} {\bf 222A} 148 

\bibitem{giorgini96}
Giorgini~S, Pitaevskii~L~P and Stringari~S 1996 {\em Phys. Rev. A}
{\bf 54} R4633

\bibitem{griffin96}
Griffin~A 1996  {\em Phys. Rev. B} {\bf 53}  9341

\bibitem{bagnato87}
Bagnato~V, Pritchard~D~E and Kleppner~D 1987  {\em Phys. Rev. A} {\bf 35}  4354

\bibitem{BaymPethick96}
Baym~G and Pethick~C 1996  {\em Phys. Rev. Lett.} {\bf 76}  6

\end{thebibliography}
\end{document}